# Fast Maximum Power Point Tracking for PV Arrays under Partial shaded Conditions

M. Amin Ghasemi, Mostafa Parniani, S. Fariborz Zarei, Hossein Mohammadian Foroushani
SHARIF UNVERSITY OF TECHNOLOGY (SUT)
Sharif university of technology, Azadi avenue,Tehran, Iran
Tel.: +98 / (21) – 6616.5963.
E-Mail: m_amingh@ee.sharif.edu, parniani@sharif.edu, zarei_fariborz@ee.sharif.edu,
hmohammadian@alum.sharif.edu
URL: http://www.sharif.ir/web/en

## Keywords

## Abstract

P-V characteristic of photovoltaic (PV) arrays under partially shaded conditions (PSC) has multiple peaks, and conventional maximum power point tracking (MPPT) algorithms may fail to track the global maximum power point (GMPP) because of their insufficient intelligence in discriminating the local and global peaks. This paper proposes a novel fast MPPT method to achieve GMPP of PV array under all PSCs. The proposed method reduces the number of required samples and increases the speed of GMPP tracking based on comprehensive study of I-V and P-V characteristics of PV array. Performance of the proposed method has been evaluated in simulations of different PSCs. Also, its performance has been compared with two selected methods in the literature through simulations. Comparisons highlight the superiority of the proposed method.

## Introduction

Photovoltaic power systems have been commercialized in many countries due to their potential long-term advantages [1]. PV arrays are the fundamental elements of PV systems, in which PV modules are connected in series and parallel. In a PV array, there is nonlinear relation between its voltage and current, and only in one operating voltage, maximum power is generated. This optimal operating point will vary as the environmental conditions change. Hence, finding this optimal point and extracting maximum power from a PV system in all environmental conditions are the main targets in this field.

To guarantee the optimal operation of large PV arrays under uniform irradiance condition (UIC), various maximum power point tracking (MPPT) techniques have been proposed in the literature. Hill climbing (HC) [2], perturb and observe (P&O) [3], and incremental conductance [4] are the most widely used methods in the literature [5]. References [6-8] have compared various MPPT methods in great details. Although these methods are quite simple in implementation, they cannot work properly in partial shading conditions (PSC). In PSC, the entire modules of an array do not receive the same solar irradiance, and P-V characteristic of array usually has multiple peaks, which include one global peak and several local peaks. PSCs are inevitable especially in PV systems installed in urban areas and in areas where low moving clouds are common [9]. If the control system cannot react to this situation, properly, the PV system will be diverted from the optimal operation mode. Given the measurements carried out under potential operating conditions of PV systems, the power loss due to MPPT failure might be up to 70% [10]. Hence, modified MPPT methods have been introduced in the previous works to overcome this big challenge.

When the PV power suddenly changes beyond a certain threshold, the proposed method in [11] detects it as PS occurrence, and starts sampling the P-V characteristic of the array with intervals of 80% of open circuit voltage of module ($V_{oc-\text{mod}}$) intervals which is supposedly the minimum interval between the peaks. Then, at each sample, when $dP/dV < 0$, there is a MPP at its vicinity, and P&O technique is called to track it. To limit the search area, it also uses another critical observation that when traversed from either side of the P–V curve, the magnitude of the peak is increasing until the global MPP (GMPP) and decreasing after that. Finally, comparing all tracked peaks, GMPP is determined. The proposed method in [12] is also based on the suggested method in [11].

Reference [13] proposes two methods. The first one samples the P-V curve and limits the search area based on short circuit current of the modules and the highest local power. The second one estimates the local MPP power by measuring the currents of bypass diodes of the modules in the array. This method improves the speed of tracking, but it increases the implementation cost.

Some papers in the literature have looked to MPPT problem as an optimization problem. The selected optimization algorithm for MPPT problem is expected to have some features like simple computational steps, fast convergence, and guaranteed convergence to GMPP together with feasibility of implementation in a low-cost digital controller [14]. Several intelligent algorithms, such as flashing fireflies [14], artificial bee colony [15], fuzzy-logic control [16], chaotic search [17], and particle swarm optimization (PSO) [18] are used to find GMPP in PSCs. Intelligent algorithms can adaptively and accurately track the GMPP regardless of shading patterns, P-V characteristic and configuration of the PV array. But they have complexity in implementation and the initial point should be selected judiciously to guarantee the convergence. Also, GMPP is obtained by numerous sampling at different points of the array P-V characteristic, which decreases the speed of tracking.

Reference [19] proposes a method that scans the PV array using ramp change of the array voltage instead of its step-like change. Since transients of the array voltage are ignorable, and the sampling interval is eliminated by using this method, this approach has acceptable speed.

Generally, a good MPPT algorithm should track the GMPP in all conditions rapidly to get high efficiency. It should also have a simple implementation with a low computational load.

Briefly explaining, the proposed method in this paper takes samples from P-V characteristic of the array at integer multiples of $V_{mpp-mod}$. Then, it determines the upper bound for the probable local peaks in the vicinity of each sample. Finally, by comparing these upper bounds and the maximum sampled power, the search area is limited, and the GMPP is obtained very quickly.

The rest of the paper is organized as follows. Sec. II discusses the characteristics of PV arrays under UIC and PSC. Sec. III proves some critical observations in PV arrays under PSC, and based on those arguments, the new MPPT method is proposed. Effectiveness of the proposed method is demonstrated in Sec. IV by simulations. Also, the proposed method is compared with selected previous approaches in this field. Finally, conclusions of the study are reported in the Sec. V.

## Characteristics of PV Array under PSC

Overall schematic diagram of a PV system is shown in Fig. 1. PV array is the main element of a PV system and consists of solar cells. DC/DC converter plays the main role in absorbing power from the PV array by controlling its voltage.

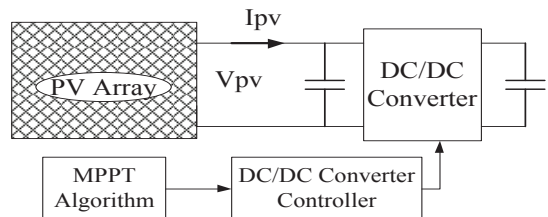

Fig. 1. Overall schematic diagram of a PV system

Among different models available in the literature for solar cells and solar modules, single-diode model, that is shown in Fig. 2, is used in this paper. Based on this model, the relation between voltage (V) and current (I) of a PV module is expressed as follows:

$$I = I_{pv} - I_o \left[\exp\left(\frac{V + R_s I}{AV_T}\right) - 1\right] - \frac{V + R_s I}{R_{sh}} \quad (1)$$

where $I_{pv}$ is the equivalent photocurrent of the module, $I_o$ is the reverse saturation current of the equivalent diode, A is the diode ideality factor, and $V_T$ is the thermal voltage of the module. Also, $R_s$ and $R_{sh}$ are the equivalent series and shunt resistances of the module.

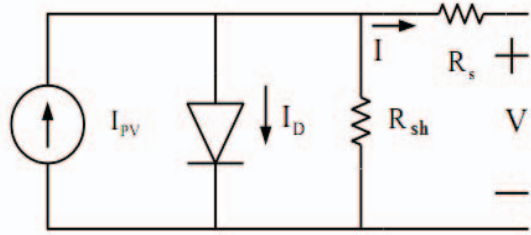

Fig. 2. Single diode electrical model of a PV module

The maximum current of module is achieved when V=0 and is known as short circuit current ($I_{sc}$). For voltages greater than open circuit voltage of module ($V_{oc-mod}$), the current will be negative, but the blocking diode, which is put in series with the module forces it to zero. To provide desired voltage and current levels, a PV array includes several parallel PV strings, in each of which there are series of several PV modules. In UIC, the maximum power points of module and array are unique and are achieved at $V_{mpp-mod} = \alpha V_{oc-mod}$ and $V_{mpp-arr} = \alpha V_{oc-arr} = N_s V_{mpp-mod}$, respectively; where $\alpha$ is a coefficient that is dependent on the module type, and usually is about 0.8, and $N_s$ is the number of series modules in each string.

When a PV string is under PSC, currents of the modules in the string are not equal. Obviously, the string current is equal to the current of the modules which receive the highest solar irradiance and generate the highest current. Therefore, portion of the string current, which is in excess of the shaded modules current, passes through their parallel resistance and generates negative voltage across them. In addition to current imbalance between the strings, hot-spots can also be resulted which damage the module. In order to protect the modules from the potential damage, a bypass diode is connected in parallel to each PV module. Although the bypass diode limits the negative voltage across the shaded module to about 0.7-1 V, it can cause multi MPP in P-V characteristic of the array and complicates the MPPT process. It is worthy to note that locations of MPPs are not specific, and depend on the shading pattern.

To discuss PSC further, assume a string with $N_s$ series modules under PSC. For the sake of simplicity, it is supposed that the string is subjected only to two different irradiance levels (S). $n_{in}$ is the number of modules that receive high irradiance level ($HS$) and are called insolated modules. $n_{sh}$ is the number of those modules which receive lower irradiance level ($LS$) and are named shaded modules. For currents higher than $I_{sc}$ of shaded modules, their bypass diodes conduct extra current and cause the voltage across them to be about $-0.7$ to $-1\,V$. In this situation, the string voltage is equally divided only between the insolated modules. For currents lower than $I_{sc}$ of shaded modules, insolated modules operate in approximately constant voltage area, and therefore, the voltage across each of these modules will be more than $V_{mpp-mod}$ and is close to $V_{oc-mod}$. Accordingly, the P-V characteristic of the string has two MPPs. The first one occurs at $V_{mpp-1} \approx n_{in} V_{mpp-mod} - n_{sh} * 0.7$ and the second MPP occurs when the voltage of one shaded module is about $V_{mpp-mod}$. In this situation, voltage of insolated modules is greater than $V_{mpp-mod}$, and therefore, the string voltage in this local MPP ($V_{mpp-2}$) is greater than $N_s V_{mpp-mod}$.

P-V characteristic of a PV array comprising several parallel PV strings, is summation of P-V characteristics of its strings. Therefore, from the above arguments two facts can be inferred from P-V characteristic of a PV array. First, the minimum difference between the voltages of two consecutive local MPPs is more than $V_{mpp-mod}$. Second, in a string and similarly in an array, the potential MPPs are located around an integer multiple of $V_{mpp-mod}$.

The second fact is used in [12], in which it takes samples from P-V characteristic of the array in integer multiples of $V_{mpp-mod}$. Then, it compares the sampled powers to determine the GMPP. In the next section, it will be shown that although Fact 2 is valid for a string, it is not true for an array and there is some deviation between the voltage of potential MPPs and integer multiples of $V_{mpp-mod}$. Therefore, the proposed method in [12] may make mistake in GMPPT. Finally, a modified version of Fact 2 is extracted and accordingly a new MPPT method is proposed. The proposed method guarantees GMPPT in all conditions.

## Proposition of the new MPPT algorithm

In this section, a modified version of the above-mentioned Fact 2 is extracted. The fact is then used in the final proposed MPPT algorithm. Discussions are presented on a sample PV array as shown in Fig. 3 in which PV modules have $V_{oc-mod} = 30V$ and $V_{mpp-mod} = 24V$ in standard condition (T=25C' and S=1kW/m²).

### Deviation Between voltage of potential MPP and integer multiples of $V_{mpp-mod}$

As explained in the previous section, in shaded PV string, the first MPP occurs at $V = n_{in}V_{mpp-mod}$. It is obvious that at MPP the following equality is valid.

$$\frac{dP}{dV} = 0 \tag{2}$$

Using (2), (3) can be obtained.

$$\frac{dI}{dV} = -\frac{I}{V} \tag{3}$$

and

$$|\frac{dI}{dV}| = |\frac{I}{V}| \tag{4}$$

Accordingly, at the points in the right side of MPP $|\frac{dI}{dV}| > |\frac{I}{V}|$ and for the points in its left side $|\frac{dI}{dV}| < |\frac{I}{V}|$. Equation (3) is the basis of IC method for MPPT.

In the following, it will be shown that the potential MPPs do not occur exactly at the integer multiples of $V_{mpp-mod}$ at all times. P-V and I-V characteristics of an array are summations of P-V and I-V characteristics of its individual strings. For the 2-string sample array, existence of MPP at around $V = mV_{mpp-mod}$ ($m$ is a positive integer number) may arise in two cases: 1) Both strings have MPP at the point $mV_{mpp-mod}$, or 2) only one of the strings (for example the first one) has MPP at $mV_{mpp-mod}$. In the first case, MPP of both strings, and therefore, MPP of the array are at $V = mV_{mpp-mod}$. In the second case, the first string has MPP at $mV_{mpp-mod}$, and therefore,

$$\left|\frac{dI_{str1}}{dV}\right|_{mV_{mpp-mod}}\right| = \left|\frac{I_{str1}}{mV_{mpp-mod}}\right| \tag{5}$$

while the second string is at constant current region at $V = mV_{mpp-mod}$ and

$$\left|\frac{dI_{str2}}{dV}\right|_{mV_{mpp-mod}}\right| < \left|\frac{I_{str2}}{mV_{mpp-mod}}\right| \tag{6}$$

Based on the above argument, it can be concluded that

$$\left|\frac{dI_{arr}}{dV}\right|_{mV_{mpp-mod}}\right| = \left|\frac{d(I_{str1}+I_{str2})}{dV}\right|_{mV_{mpp-mod}}\right| < \left|\frac{I_{str1}+I_{str2}}{mV_{mpp-mod}}\right| = \left|\frac{I_{arr}}{mV_{mpp-mod}}\right| \tag{7}$$

Therefore, the prospective MPP of the array is at the right side of $V = mV_{mpp-mod}$. Indeed, at the voltages greater than $mV_{mpp-mod}$, $|\frac{dI_{str1}}{dV}|$ increases rapidly and compensates the greater value of $\frac{dI_{str2}}{dV}|_{mV_{mpp-mod}}$ in (7). Hence, the equality (4) holds at $V > mV_{mpp-mod}$.

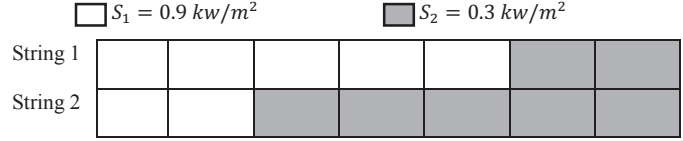

Fig. 3. the sample array under a specific PS pattern.

In Fig. 4 I-V and P-V characteristics of String 1, String 2, and the array have been shown. Shading patterns of the corresponding array are shown in Fig. 3. At $V = 5*24\ v$, the first string has MPP and the second string is at constant current region. Hence, MPP of the array has occurred at $V > 5*24$. The same situation holds for the array at $V = 2*24\ v$.

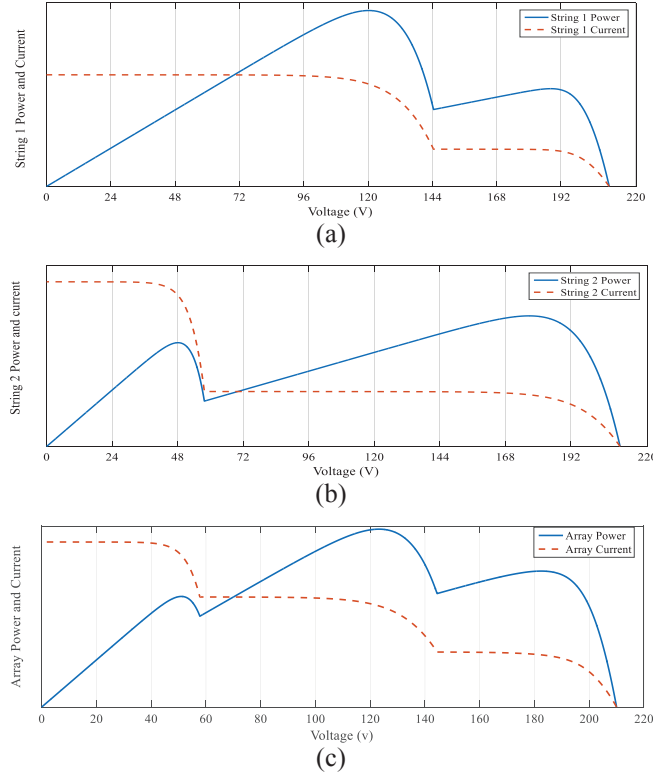

Fig. 4. I-V and P-V characteristics of String 1 (a), String 2 (b), and the array (c).

In the above discussion, it was shown that the potential MPPs of an array occur at $V_{mpp} \geq mV_{mpp-mod}$. In the following, the maximum difference between $V_{mpp}$ and the sampled voltage $mV_{mpp-mod}$ will be determined. For a string that has MPP at $V = mV_{mpp-mod}$, $\left|\frac{dI_{str}}{dV}\right| \approx 0$ for $V > mV_{oc-mod}$ (see Fig. 5 for m=2), and because of non-zero value for $I_{arr}$ at this voltage, according to (4), the potential MPP of array cannot occur at $V > mV_{oc-mod}$. Therefore, the maximum difference between the sample and the potential MPP cannot be greater than $mV_{oc-mod} - mV_{mpp-mod}$ (Fig. 5). Furthermore, since sampling from P-V characteristic of the array is done at the intervals $V_{mpp-mod}$, the maximum difference between each MPP of the array and the nearest back sample is lower than $V_{mpp-mod}$ (Default Limit in Fig. 5). Therefore, the maximum difference between each MPP of the array and the nearest back sample is as follows.

$$\Delta V_{max-m} = \min\{V_{mpp-mod}\ \&\ m(V_{oc-mod} - V_{mpp-mod})\} \qquad (8)$$

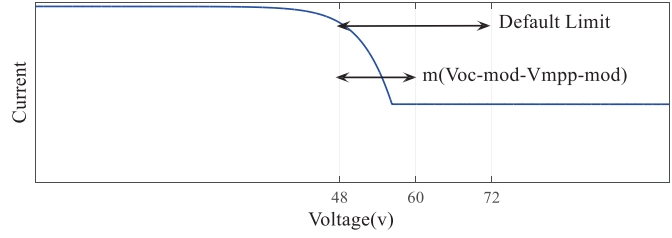

Fig. 5. Maximum deviation between MPP and the samples from integer multiples of $V_{mpp-mod}$ for $m = 2$ ($V_{oc-mod} = 30$ and $V_{mpp-mod} = 24$).

**Deviation between the sampled power at integer multiples of $V_{mpp-mod}$ and power of potential MPP**

In the previous section, the maximum possible deviation between the potential MPP of the array and samples from integer multiples of $V_{mpp-mod}$ was determined. This voltage deviation yields a corresponding difference between the sampled power $P_{s-m}$ and the potential MPP power. In this section, the maximum difference between the sampled power and power at the potential MPP of array is specified. To reach this goal, the I-V characteristic of the array is approximated with step-wise characteristic as shown in Fig. 6, in which the width of each step is $\Delta V_{max-m}$.

It is clear that with increasing the voltage of the array its current decreases. Therefore, the step-wise approximation of I-V curve is always above the real I-V curve. Hence, active power resulting from this approximate I-V curve at the corner points are the maximum probable power of array at the limit point voltages. These power limits are calculated using the sampled powers ($P_{s-m}$) and $\Delta V_{max-m}$ as follows.

$$P_{up-m} = P_{s-m} + I_{s-m}\Delta V_{max-m} \tag{9}$$

where $I_{s-m}$ is the array current at the m'th voltage sample ($V = mV_{oc-mod}$).

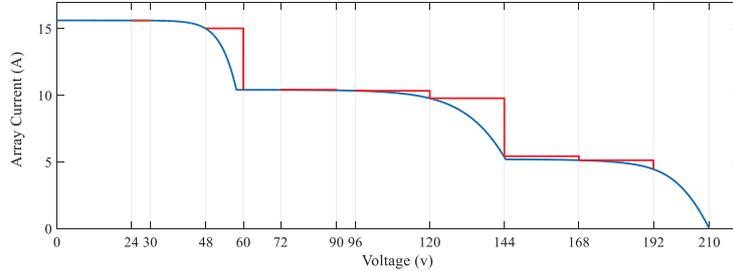

Fig. 6 Step-wise approximation of PV array I-V characteristic.

**Proposed MPPT algorithm**

Based on the above discussions, the proposed MPPT method is presented as follows.

1) Upon detecting a PSC using the proposed method in [19], sampling from P-V characteristic of the array is done at integer multiples of $V_{mpp-mod} \approx 0.8 V_{oc-mod}$ in the following voltage range, which is the range that MPPs can take place [13].

$$V_{mpp-mod} < V < 0.9 V_{oc-arr} \tag{10}$$

2) Considering (8) and the step-wise approximation illustrated in Fig. 6, for each sample, the voltage range $[mV_{mpp-mod} \quad mV_{mpp-mod} + \Delta V_{max-m}]$ is the region in which probable MPP can take place. Also, the maximum probable values of the array power ($P_{up-m}$) at each voltage range are calculated using (9).

3) The maximum power point among samples is selected and operating voltage of the array is set to the corresponding voltage. Then, P&O algorithm is called to track the exact local MPP at its vicinity. PV power at this point is recorded as $P_{mpp}$. According to the presented arguments in Sec.

III-A, since MPP takes place at $V \geq mV_{mpp-mod}$, in $P_{mpp}$ tracking process, voltage of the array may only increase.

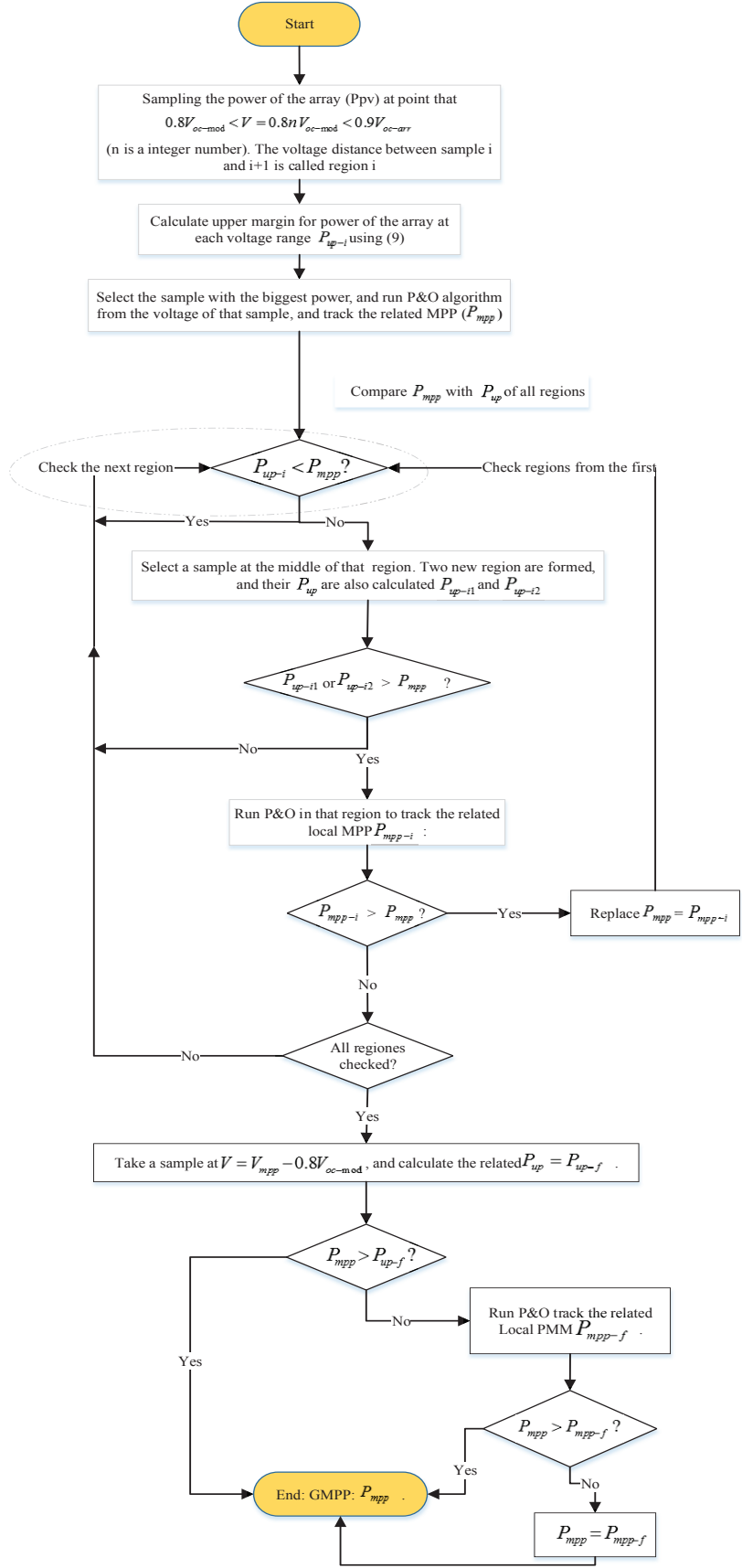

Fig. 7. Complete flowchart of the proposed algorithm.

4) $P_{mpp}$ is compared with all $P_{up-m}$ values calculated in the voltage region (10), except to two neighboring $P_{up}$s. Indeed, right side region is searched once to track $P_{mpp}$, and therefore, the corresponding $P_{up}$ is not considered in the mentioned comparison (The exception about left side neighboring $P_{up}$ will be investigated in Step 5.). If any $P_{up-m}$ is greater than $P_{mpp}$, one obvious way is that the array voltage is set to the corresponding $mV_{mpp-mod}$, and P&O algorithm is called to track the real local MPP at that vicinity. Then, it is compared with $P_{mpp}$. However, in order to reduce the number of P&O searches, is further refined as follows. At first, $\Delta V_{max-m}$ is divided to two equal section, and the array power is measured at its middle point $m_1$. Calling this sample power $P_{s-m1}$, the corner point powers $P_{up-m1}$ and $P_{up-m2}$ are recalculated as follows.

$$P_{up-m1} = P_{s-m} + I_{s-m} \cdot \Delta V_{max-m/2}$$
$$P_{up-m2} = P_{s-m1} + I_{s-m1} \cdot \Delta V_{max-m/2} \quad (11)$$

Then, $P_{mpp}$ is compared with $P_{up\_m1}$ and $P_{up-m2}$, If each of these two powers is greater than $P_{mpp}$, P&O algorithm is called to track the exact MPP at that vicinity. If the new MPP has greater power than $P_{mpp}$, $P_{mpp}$ is replaced with that power, and this step (step 4) is repeated.

5) Step 4 is repeated until there is no voltage region with $P_{up}$ greater than $P_{mpp}$, except the one at the left side of $P_{mpp}$, as mentioned earlier. According to the presented arguments in Sec. II, there could be no other MPP at the interval

$$[V_{mpp} - 0.8V_{oc-mod} \quad V_{mpp}] \quad (12)$$

Therefore, $P_{mpp}$ is compared with the maximum probable value of array power at the voltage region between $V_{mpp} - 0.8V_{oc-mod}$ and the nearest back sample, as defined in (9). This value is calculated and recorded as $P_{up-f}$. If it was lower than $P_{mpp}$, GMPP has been already achieved; else, there is another peak at that vicinity and P&O algorithm is called to track it. Comparing the newly founded peak with $P_{mpp}$ yields the GMPP.

6) Now, operating voltage of the array is taken to that point and P&O algorithm is called to continue local MPPT.

Complete flow chart of the proposed method is shown in Fig. 7.

## Simulation Results

In this section, using different simulations, performance of the proposed method in tracking of GMPP under PSC is evaluated. The simulated system configuration is as shown in Fig. 1. To examine the performance of the suggested method, the results are compared with two other methods that are available in [11] and [13].

The method proposed in [13] considers voltage range of $[0.8V_{oc-mod} \quad 0.9V_{oc-arr}]$ as search region for GMPP and reaches to GMPP based on three processes: searching process, skipping process, and judging process, which has been called search-skip-judge global MPPT(SSJ-GMPPT) method [13]. Description of the method presented in [11] is given in Sec. I.

For better analysis and comparison, two different arrays with given partially shading patterns are simulated. The first array is the one shown in Fig. 3. At first, the array is under UIC with irradiance level $S_1 = 0.9 \ kw/m^2$. During UIC, P&O algorithm tracks MPP. Then, the PSC occurs at t = 0.3s and irradiance level for shaded modules is $S_2 = 0.3 \ kw/m^2$. The second array is a $3X7$ array shown in Fig. 8 with the same modules of the first array. It is first under UIC, with irradiance level $S_1 = 0.9 \ kw/m^2$, and then goes under PSC at t = 0.3s, which irradiance level for shaded modules is $S_2 = 0.35 \ kw/m^2$.

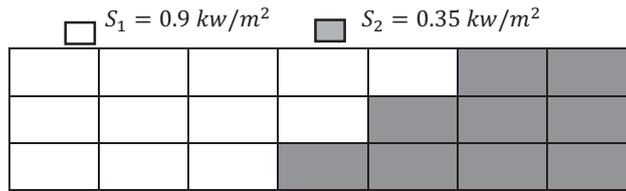

Fig. 8. Shading pattern of the second sample array.

P-V and I-V characteristics of the second array under the given PSC are shown in Fig. 9.

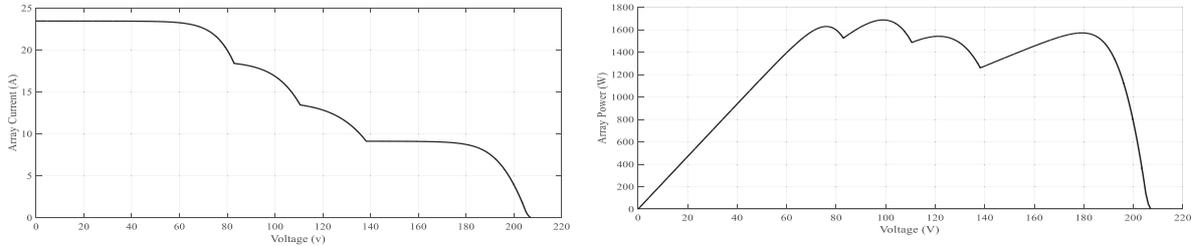

Fig. 9. I-V and P-V characteristics of the second simulated Array.

Voltage increment in the search process for all algorithms is considered 1V and the voltage search region is as (10). Furthermore, to ensure that the system attains steady state before the next MPPT perturbation is initiated, the sampling interval is chosen as 20ms.

Simulation results of GMPP tracking processes for the first array, using the three methods are presented in Fig. 10.

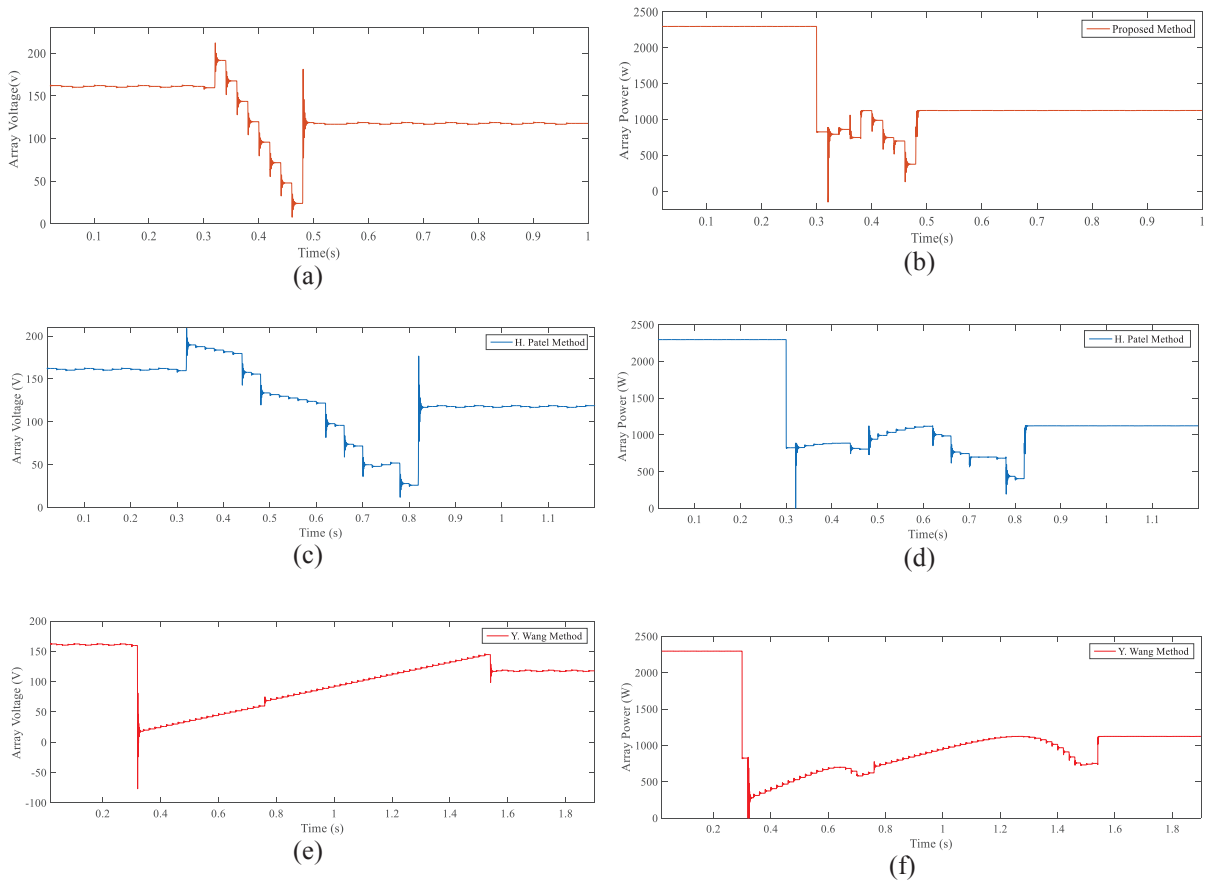

Fig. 10. GMPP Tracking process of the three methods under the given shading patterns for the first array, (a, b) The proposed method, (c, d) presented method in [11], and (e, f) presented method in [13].

Numerical results of GMPPT using the proposed method is also presented in Table I.

Maximum power of all samples is $P = 1175W$ at $V = 5 * 24 = 120V$. Hence, P&O is called to track exact value of the local MPP. As it was expected, local MPP occurs at $V_{mpp} = 123V > 120V$ and $P_{mpp} = 1179W$. Maximum probable power at none of the voltage regions is greater than $P = 1179W$, except the region at left side of $V_{mpp}$, i.e. [96 120]. Therefore, no further search is needed. Then, $P_{up-f}$ at the left side of $V_{mpp}$ is calculated at $V = V_{mpp} - 24 = 123 - 24 = 99V$. This maximum probable power of the interval [96V  99V] is $P_{up-f} = 1025W$, which is lower than $P_{mpp} = 1179$. Therefore, GMPP tracking terminates at $V = 123V$. From this moment, P&O algorithm is called to track small changes of MPP.

With the proposed method, total GMPP process needs 12 samples and lasts $240ms$. In contrast, the method presented in [11] needs 25 samples and lasts about $500ms$. Also, the presented method in [13] needs 62 samples and lasts about $1250ms$. These results show that all methods are able to track the GMPP. However, the presented method in this paper is considerably faster than the other two methods. Results of the three mentioned methods for the second array are shown in Fig. 11.

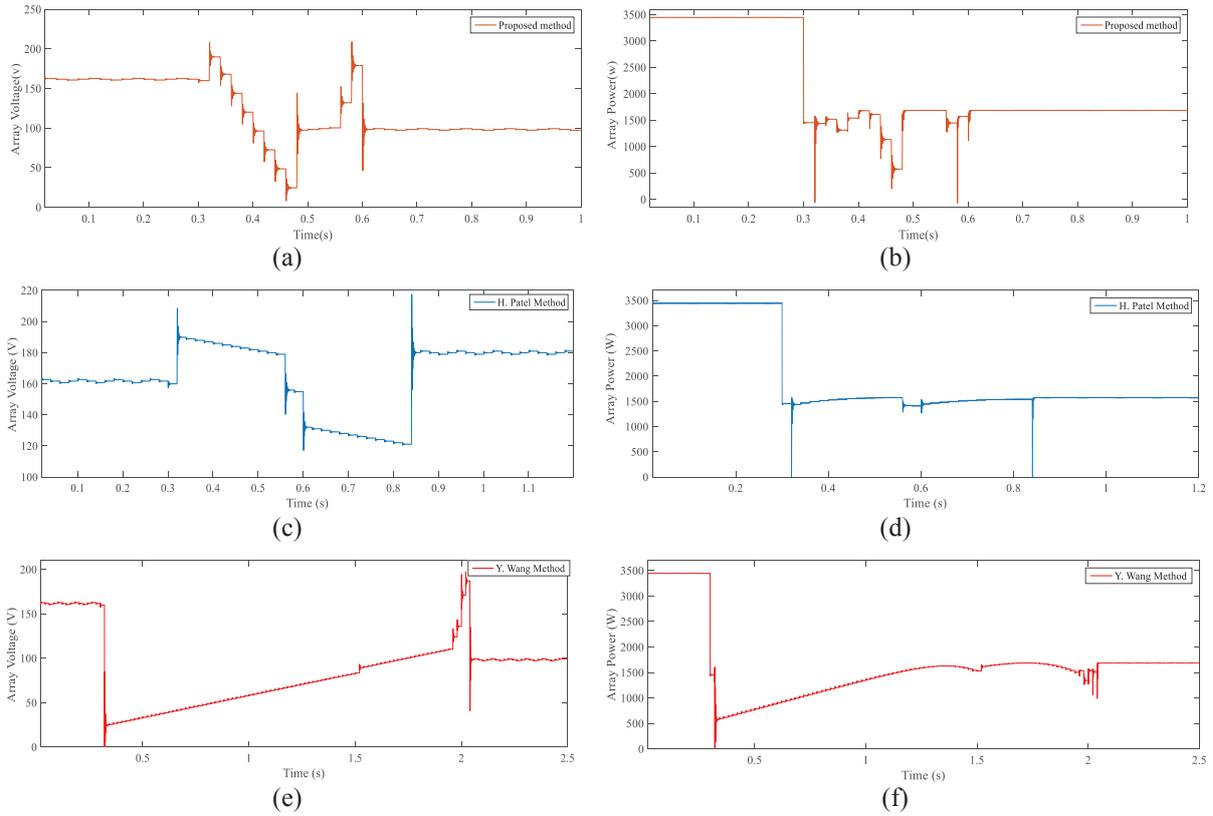

Fig. 11. GMPP tracking process of the three methods under the specified shading pattern for the second array, (a, b) The proposed method, (c, d) presented method in [11], and (e, f) presented method in [13].

Similar to the first simulation, the process of GMPPT for the second array using the proposed method is also presented in Table II.

The maximum power sample is $P = 1680W$ at $V = 4 * 24 = 96V$. Hence P&O is called to track exact value of the corresponding local MPP. As expected, the local MPP occurs at $V_{mpp} = 99V > 96V$ with $P_{mpp} = 1688W$. Maximum probable power $P_{up}$ at the voltage region [120 144] is greater than $P_{mpp}$,(1836 > 1688). Therefore, another sample is taken at middle of this region (i.e. $V = 132V$). Measured power of this point is $1445W$, the maximum probable power in the region [120  132] is $P_{upi-1} = 1683W$ and in the region [132 144] is $P_{upi-2} = 1570W$. Both of them are lower than $P_{mpp} = 1688W$ meaning that there is no MPP with more power than $P_{mpp} = 1688W$ in the region [120  144]. Similarly, maximum probable power $P_{up}$ in the voltage region [168 190] is greater than $P_{mpp}$,(1720 > 1688). Therefore, a sample at the midpoint $V = 179V$ is taken. Its power is $P = 1571W$, the maximum probable power in the region [168  179] is $P_{upi-1} = 1620W$ and in the region [179  190] is $P_{upi-2} =$

$1667W$. Again, both of them are lower than $P_{mpp} = 1688W$, and there is no MPP with more power than $P_{mpp} = 1688W$ within the range [168  190].

**TABLE I: Process of GMPPT using the proposed method for the first array.**

| $V_s$ (V) | $P_s$ (W) | Voltage regions | $P_{up}$ (W) |
|---|---|---|---|
| 192 | 858 | 190-168 | 987 at V=190V |
| 168 | 864 | 168-144 | 900 at V=168V |
| 144 | 772 | 144-120 | 1410 at V=144V |
| 120 | **1175** | 120-96 | 1242 at V=120V |
| 96 | 994 | 96-72 | 937 at V=90V |
| 72 | 750 | 72-48 | 900 at V=60V |
| 48 | 720 | 48-24 | 471 at V=30V |
| 24 | 377 | | |
| 121 | Run P&O to reach P=1179W at V=123 | 120-144 | 1179, This is the exact value achieved by P&O |
| 122 | | | |
| 123 | | | |
| 124 | | | |

**TABLE II: Process of GMPPT using the proposed method for the second array.**

| $V_s$ (V) | $P_s$ (W) | Voltage regions | $P_{up}$ (W) |
|---|---|---|---|
| 192 | 1447 | 190-168 | 1720 |
| 168 | 1521 | 168-144 | 1531 |
| 144 | 1313 | 144-120 | 1836 |
| 120 | 1530 | 120-96 | 2100 |
| 96 | **1680** | 96-72 | 2007 at V=90V |
| 72 | 1606 | 72-48 | 1406 at V=60V |
| 48 | 1125 | 48-24 | 703 at V=30V |
| 24 | 563 | | |
| 97 | Run P&O to reach P=1688W at V=99 | 96-120 | 1688, This is the exact value achieved by P&O |
| 98 | | | |
| 99 | | | |
| 100 | | | |
| 132 | 1445 | 120-132 | 1683 |
| | | 132-144 | 1570 |
| 179 | 1571 | 168-179 | 1620 |
| | | 179-190 | 1667 |

To this point, all voltage regions have been analyzed, except the region at left side of $V_{mpp}$ ([72 96]). Then the maximum probable power $P_{up-f}$ at left side region of $V_{mpp}$ is evaluated at $V = V_{mpp} - 24 = 99 - 24 = 75V$. This $P_{up}$ of the region [72V  75V] is $P_{up-f} = 1672W$ which is lower than $P_{mpp} = 1688$. Therefore, GMPP tracking terminates, and GMPP will be reached at $V = 99V$. From this moment, P&O algorithm is called at $V = 99V$ to track small changes of MPP.

In this case, the proposed GMPP process needs 14 samples and lasts $280ms$. While using the presented method in [11], the first two MPPs at $V = 180V$ and $V = 122V$ are tracked after 27 samples and about $540ms$ ($P_{mpp-180v} = 1571W, P_{mpp-122v} = 1541W$). Since $P_{mpp-180v} > P_{mpp-122v}$, it terminates tracking and considers $P_{mpp-180v}$ as GMPP. However, the actual GMMP is $P_{mpp} = 1688W$ and occurs at $V = 99V$. Therefore, this method mistakes in GMPP tracking in this PSC for the given array. The presented method in [13] needs 86 samples and takes $1720ms$ to track GMPP. Simulation results and comparison of the methods show that the presented method in this paper not only tracks GMPP successfully in all conditions, but also is considerably faster than the other two methods.

## Conclusion

In this paper, I-V and P-V characteristics of PV arrays under PSC have been analyzed in detail. The maximum probable power of array in a voltage region between each two subsequence samples is calculated. Accordingly, a new fast MPP tracking method is proposed that can track GMPP of partially shaded PV array very fast, using few samples of array. It was shown that the proposed method tracks the GMPP at all PS patterns. Performance of the proposed method was evaluated through simulations. The results of this work were compared with the results of two best-claimed methods in the literature, and superiority of the suggested method over them was demonstrated.